\documentclass{aa}
\usepackage{graphicx}

\newcommand{\HI}{{\ion{H}{1}}}

\newcommand{\abHI}{\rm H{\hskip 0.05cm\scriptsize I}}

\newcommand{\kms}{$\,$km$\,$s$^{-1}$}

\newcommand{\mJybeam}{mJy beam$^{-1}$}
\newcommand{\msun}{{$M_\odot$}}

\def\HI{H{\,\small I}}

\newcommand{\ltsima} {$\; \buildrel < \over \sim \;$}
\newcommand{\gtsima} {$\; \buildrel > \over \sim \;$}
\newcommand{\lta} {\lower.5ex\hbox{\ltsima}}
\newcommand{\gta} {\lower.5ex\hbox{\gtsima}}

\input{psfig}

\begin{document}

\title{Anomalous H{\Large\hskip0.1cm I} kinematics  in Centaurus A: 
evidence for jet-induced star formation}

\titlerunning{Anomalous \abHI\ kinematics  in Centaurus A }
\authorrunning{Oosterloo \& Morganti}

\author{Thomas A. Oosterloo\inst{1} \& Raffaella Morganti\inst{1}}

\offprints{oosterloo@astron.nl}

\institute{Netherlands Foundation for Research in Astronomy, Postbus 2,
NL-7990 AA, Dwingeloo, The Netherlands}

\date{Received ...; accepted ...}

\abstract{We present new 21-cm \abHI\ observations performed with ATCA of the
large \abHI\ filament located about 15 kpc NE from the centre of Centaurus~A
and discovered by Schiminovich et al.\ (1994).  This \abHI\ cloud is situated
(in projection) near the radio jet of Centaurus A, as well as near a large
filament of ionised gas of high excitation and turbulent velocities and near
regions with young stars.  The higher velocity and spatial resolution of the
new data reveals that, apart from the smooth velocity gradient corresponding
to the overall rotation of the cloud around Centaurus A, \abHI\ with anomalous
velocities of about 100\kms\ is present at the southern tip of this cloud. 
This is interpreted as evidence for an ongoing interaction between the radio
jet and the \abHI\ cloud.  Gas stripped from the \abHI\ cloud gives rise to
the large filament of ionised gas and the star formation regions that are
found downstream from the location of the interaction.  The implied flow
velocities are very similar to the observed anomalous \abHI\ velocities. 
Given the amount of \abHI\ with anomalous kinematics and the current star
formation rate, the efficiency of jet-induced star formation is  at most
of the order of a percent. 
}

\maketitle

\section{Introduction}

Interaction between the non-thermal plasma ejected from the active nucleus and
the interstellar medium (ISM) of a galaxy is responsible for a variety of
phenomena in radio galaxies such as ionisation of the gas and AGN driven
outflows. Such interactions are considered to be particularly relevant in high
redshift radio galaxies, as they are typically living in a gas-rich
environments (see e.g.\ van Breugel 2000 and references therein). One aspect
of jet-ISM interaction is that it can trigger star formation. Such jet-induced
star formation is considered a possible mechanism to explain the UV continuum
emission observed in the host galaxies of distant radio sources and the
"alignment effect" between the radio emission and this continuum (Rees 1989).

Detecting and studying star formation produced by this mechanism in high-$z$
radio galaxies is very challenging.  The only case  where this
has been done is 4C~41.17 (Dey et al.\ 1997).  Because of the observational
problems for high-redshift sources, it is important to find nearby examples of
star formation triggered by the radio jet that can be studied in more detail.

Moreover, recent numerical simulations have shown some interesting results. It
is found that the interaction of a radio jet with a clumpy gaseous medium can
produce fragmented clouds that cool and condense very quickly (Mellema et al.\
2002, Fragile et al.\ 2004). Even for moderate jet velocities, the interaction
and the consequent production of shocks by the radio jet has the potential
to trigger large-scale star formation in a galaxy (Fragile et al.\ 2004, van
Breugel et al.\ 2003).  More data, in particular for nearby objects, are
needed to better constrain these models.

Only a few cases are known of nearby radio galaxies where off-nucleus young
stars are found and where a possible relation exists with the radio jet
(Minkowsky's object and 3C285; van Breugel et al.\ 1985, van Breugel \& Dey
1993). The best example is perhaps the nearby radio galaxy Centaurus~A.  About
15 kpc NE from the centre of the galaxy, well outside the main optical body,
groups of young stars (with an estimated age of about 15 Myr) are found at the
location where the large-scale radio jet passes large filaments of highly
ionised gas with turbulent kinematics and where also a large cloud of neutral
gas is found (Schiminovich et al.\ 1994; SGHK). The presence of these young
stars has been explained by several authors as resulting from star formation
triggered by gas shocked by the passage of the radio jet. Given the proximity
of Centaurus A, many aspects of the jet-cloud interaction, the star formation and
the spatial variation of the properties of the gas and the stars, can be
studied  much btter than in any other source and it can serve as a
prototype case.  In this paper we investigate the jet-induced star formation
hypothesis in more detail using the kinematics of the neutral hydrogen found
in this location.

\section{The north-east region of Centaurus~A}
 
The north-east region of Centaurus~A, up to a few tens of kpc from its
nucleus, is a particularly complex site where different structures have been
found and studied: 1) a complex system of filaments of ionised gas: the
so-called {\sl inner filament}, located $\sim 8$ kpc\footnote{We have assumed
a distance of Centaurus~A of 3.7 Mpc  (Hui et al.\ 1993), 1 arcsec $\sim 18$
pc.}, and the {\sl outer filament} $\sim 15$ kpc from the centre (Blanco et
al.\ 1975, Morganti et al.\ 1991); 2) a large-scale radio jet that connects
the bright inner radio lobes to the much  larger Northern Middle Lobe (NML) and
outer radio lobe (Morganti et al.\ 1999). This radio jet is either a classic
jet driven by the AGN in Cen A, or it is part of a buoyant bubble of plasma
deposited by an intermittently active jet (Saxton, Sutherland \& Bicknell
2001); 3) regions of recent star formation (Graham 1998, Fassett \& Graham
2000, Mould et al.\ 2000, Rejkuba et al.\ 2002) and 4) large \HI\ clouds that
appear to form a partial ring rotating around the galaxy (SGHK).  Near the
location of the peak of the \HI\ emission CO and cold dust are detected
(Charmandaris, Combes \& van der Hulst 2000, Stickel et al.\ 2004).

 Although the inner and outer filaments are not ionised by star formation
(Morganti et al.\ 1991, Sutherland et al.\ 1993), young stars (as young as
15~Myr) and a few small star forming regions have been detected in optical
observations at the location of both the inner and the outer filaments of
ionised gas (Mould et al.\ 2000, Rejkuba et al.\ 2002), while also recent
GALEX data (Neff et al.\ 2003) indicate the presence of young stars. In the
outer filament, the optical observations show that the stars are found in two
groups, one aligned with the radio jet and one in a north-south alignment
along the edge of the
\HI\ cloud (Rejkuba et al.\ 2002; see Fig.\ 3). This morphology is confirmed
by the GALEX data.

The large-scale radio jet appears to pass (at least in projection) very close
to the location of the \HI\ cloud as well as close to the outer filament of
ionised gas. This is illustrated in Fig.\ 1. The spatial coincidence of these
structures, together with the fact that the ionised and neutral gas have the
same overall velocity, has led to the suggestion (e.g.\ Graham 1998) that the
radio jet is interacting with the \HI\ cloud leading to the filaments of very
turbulent and highly excited ionised gas and triggering the star formation in
this region. 

If the radio jet is indeed interacting with the \HI\ cloud, one may expect to
see direct evidence in the kinematics of the \HI\ for this interaction in the
form of deviations, driven by the radio plasma, from the overall rotation of
the \HI\ cloud about the galaxy.  Due to the limited spatial and spectral
resolution of the data of SGHK, only the overall coincidence in space and
velocity of the neutral and ionised gas could be established. To investigate
whether there is indeed direct kinematical evidence for a jet-cloud
interaction, where this is occurring and how large the kinematical anomalies
are, we have obtained new \HI\ data with higher spatial and spectral
resolution.

\section{Observations and Data Reduction}

The observations were done using the Australia Telescope Compact Array (ATCA)
in two standard 750-m and 1.5-km configurations.  The observations (12 h in
each configuration) were carried out on the 12th and 20th April 2000.  The
choice of these two configurations was made to obtain good spatial resolution
for the region of interest (as imaged by SGHK). Another important
consideration was to use configurations that do not contain (many) short
baselines. Given the very strong continuum emission of the radio source in
Centaurus A, bandpass calibration is in practise impossible on short
baselines. The shortest baseline in the combination of the two configurations
is 30 m long, but the remaining ones are all longer than 100 m. The continuum
flux on the 30-m baseline was too strong to allow proper bandpass calibration
of the data on this baseline (see below) and the data from this baseline was
not used in making the images. This implies that structures in a single
channel that are larger than about 7 arcmin are not detected in our data. The
extent of the \HI\ cloud in a single channel is well below this size,
therefore not using baselines shorter than 100 m does not influence the
results.

One pointing was observed, centred on $\alpha$(J2000) = $13^{\rm h}26^{\rm
m}15^{\rm s}$ and $\delta$(J2000) =
$-42^\circ45^\prime00^{\prime\prime}$. Using this pointing centre, the strong
continuum emission from the central region of Centaurus~A was located at the
half-power point of the primary beam.  This somewhat attenuates the strong
continuum signal that otherwise would make the detection of the relatively
weak \HI\ emission more difficult.  A band of 16~MHz with 512 channels was
used and the central frequency was set at 1418~MHz, corresponding to the
velocity of the previously detected \HI\ emission ($\sim 300$ \kms, SGHK). The
final velocity resolution is $13.2$
\kms\ after Hanning smoothing the data. This is about a factor three better
than that of the data of SGHK.

The source PKS~1934--638 was used as bandpass calibrator and observed for
about 1~h. Short observations of about 5 minutes were done every hour on the
phase calibrator PKS 1315--46.  The data reduction was carried out using the
MIRIAD package (Sault, Teuben \& Write 1995).

\begin{figure*}
\centerline{\psfig{figure=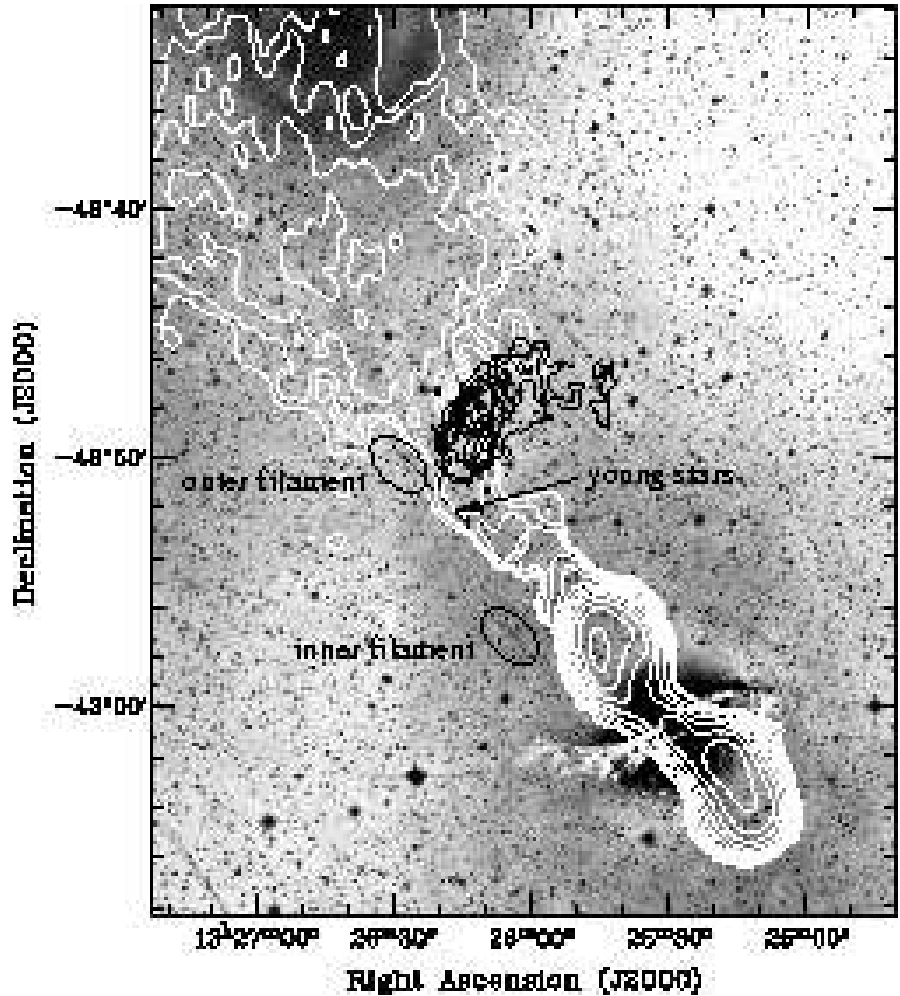,angle=0,width=13cm}}
\caption{Overlay showing the positions of the various 
components described in the text.  The optical image (kindly provided by
D.\ Malin) shows the well-known dust lane of Centaurus A and the faint diffuse optical
emission that extends to very large radius. The white contours denote the
radio continuum emission (after Morganti et al.\ 1999) showing the bright
inner radio lobes and the large-scale jet that connects these lobes to the
so-called Northern Middle Lobe (NML) in the top-left of the figure.  The black
contours denote the \HI\ cloud discussed in this paper (see also Fig.\ 2). The
locations of the inner and outer filaments of highly ionised gas are
indicated, as well as the location of young stars (see also Fig.\ 4).}
\end{figure*}

As also described by SGHK, the extremely strong radio continuum of Centaurus~A
($> 100$ Jy on the shortest baseline) makes   the bandpass calibration 
the most delicate part of the data reduction and it requires extra care to
make sure that fairly flat spectral baselines are obtained.  Even the strong
bandpass calibrator PKS~1934--638 (14.9 Jy at the observed frequency) is not
strong enough compared to Centaurus~A. In order to minimise the contribution
of extra-noise from the bandpass calibration, we have smoothed the bandpass
calibration obtained from PKS~1934-638 with a box-car filter 15 channels
wide. This effectively increases the flux level of PKS 1934--638 to about 60
Jy which is higher than the detected flux on almost all baselines of Centaurus
A. This smoothed bandpass correction has been applied to the {\sl unsmoothed}
data of Centaurus~A. This ensures that the velocity resolution is preserved
without increasing too much the noise of the data. This procedure to calibrate
the bandpass appears to work very well, mainly due to the relatively flat
intrinsic bandpass-shape that is typical of the ATCA.  Only for the data on
shortest (i.e.\ 30-m) baseline it failed to produce usable spectra and for
this reason the data of this baseline were not used in making the line-images
of the \HI\ cloud.  Only residuals that are very broad in frequency remain in
the data, but this does not have any effect on the conclusions of this
paper. The noise level in the final datacube is 1.5 \mJybeam\ while the
theoretical noise level for the instrumental setup used is 1.1 \mJybeam. Hence
our method of calibrating the spectral response appears to have worked fairly
well.

In order to improve on the phase calibration obtained from PKS 1315--46,
frequency independent selfcalibration was also performed taking advantage of
the strong unresolved nuclear \HI\ absorption, still visible in a few channels
despite the large offset from the nucleus of our pointing centre.

The subtraction of the continuum was done using the task UVLIN, which
makes a linear fit to the continuum in the visibility data and
subtracts it.

The final cube was made using the combined dataset using robust weighting
 (Briggs, 1995) with the robustness set to 0.  The restoring beam is $29.1
\times 19.5$ arcsec in ${\rm PA} =-4.7^\circ$.  The spatial resolution of
these observations is therefore slightly more than a factor two higher than
what obtained by SGHK.  The 3-$\sigma$ detection limit over 13.2
\kms\ is $1.0\times10^{20}$ cm$^{-2}$ which is the same as the detection limit
in the SGHK data.

In our data, we detect mainly the large \HI\ cloud that peaks 15$^\prime$ NE
of the nucleus. The other \HI\ detected by SGHK that is part
of the outer \HI\ ring, as well as the \HI\ associated with the well-known dust
lane of Centaurus A, is visible in our data, but the signal-to-noise of this
emission is strongly affected by  the primary beam.  Therefore,
we will not discuss this further as it is much better studied by the
SGHK observations.

\begin{figure}
\centerline{\psfig{figure=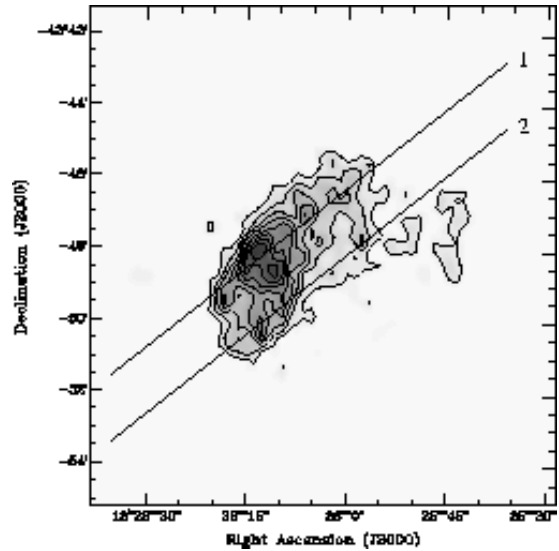,angle=0,width=8cm}}
\caption{Total HI intensity image of the NE \HI\ cloud. The locations of the
two position-velocity plots of Fig.\ 3 are indicated.  Contour levels are 1, 3, 5, 7,
9, 11, 13, $15\times10^{20}$ cm$^{-2}$}
\end{figure}

\section{Results: morphology and kinematics of the \HI\ cloud.}

\subsection{Morphology of the \HI\ cloud}

The total intensity image of the NE \HI\ cloud is shown in Fig.\ 2, while in
Figs 1 and 4 the position of the  \HI\ cloud can be seen in relation to the
large-scale radio jet, the optical filaments and the young stars.

Except for the higher spatial resolution, the morphology of the \HI\ agrees
well  with that obtained by SGHK.  The extent of the cloud is
about $8^\prime\times3^\prime$, corresponding to about $8.6\times 3.2$ kpc.
The peak column density is $1.7 \times 10^{21}$ cm$^{-2}$, which is perfectly
consistent (considering the difference in spatial resolution) with the value
$1 \times 10^{21}$ cm$^{-2}$ of SGHK.  The total \HI\ mass of the cloud is
$5.6\times10^7$ \msun.  SGHK give a total mass of $1.5\times10^8$ \msun\ for
all \HI\ clouds at large radius combined.  Given the relative sizes of the
clouds detected by SGHK, our mass estimate  appears well consistent with their
results.  This underlines that the missing short baselines in our observations
have not affected our images.

\begin{figure}
\centerline{\psfig{figure=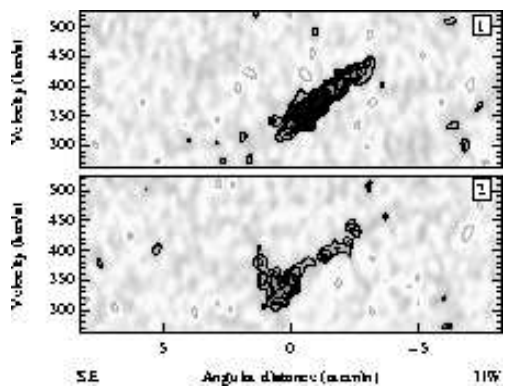,angle=270,width=8.5cm}}
\caption{Position-velocity plots taken along the lines indicated in Fig.\ 2.
 Contour levels are --6.75, --4.5, 4.5 ($3\sigma$), 6.75, 9.0... \mJybeam.
}
\end{figure}

\subsection{Kinematics of the \HI}

While the morphology of the  \HI\ emission agrees  with the results
presented by SGHK, it is the kinematics of the \HI\ in the new observations
that gives the interesting new result. Due to the higher velocity resolution
of the new data we can investigate this in much more detail compared to what
previously done by SGHK.

The main characteristic of the kinematics of the gas is illustrated by the two
position-velocity slices presented in Fig.\ 3. They are taken along position
angle $-42^\circ$ centred on two different positions of the cloud (as
indicated in Fig.\ 2). The slices show, as the data of SGHK, that the overall
kinematics of the cloud is characterised by a smooth velocity gradient of
about 30\kms\ kpc$^{-1}$ corresponding to the rotation of the cloud around the
galaxy.  However, the slice taken along line 2 clearly shows that the most
southernly region along this slice (corresponding to the southern tip of the
cloud) does not conform to this regular gradient and that it shows a velocity
anomaly. The southern tip of the cloud shows, when going to the SE, an abrupt
reversal of the large-scale velocity gradient in the form of a sudden upturn
in the velocities of about +100\kms\ over 1 kpc. This upturn is not observed
in the parallel slice centred on a slightly more northernly position. The
anomalous velocities are therefore confined to a small region (Fig.\ 4).

\begin{figure*}
\centerline{\psfig{figure=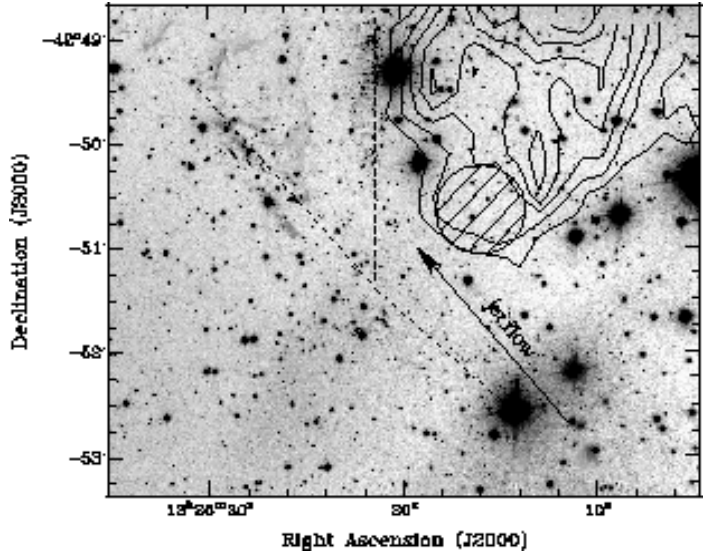,angle=270,width=12cm}}
\caption{\HI\ contours of the southern region of the \HI\ cloud, 
drawn on top of a broad band optical image (kindly provided by
M.\ Rejkuba). The hatched area indicates the location where the anomalous \HI\
velocities are detected while the arrow indicates the location and the flow
direction of the radio jet. The filament of ionised gas is visible in the top
left.  The dashed lines roughly indicate the locations of young stars. Contour
levels are as in Fig.\ 2}
\end{figure*}

\section{Discussion}
\subsection{Is the velocity anomaly indicating a jet-cloud interaction?}

The central question is whether the velocity anomaly observed at the S tip of
the \HI\ cloud is the result of an interaction with the radio jet. In
principle, the reversal of the velocity gradient could simply be reflecting
the overall kinematics (driven by gravity) of the cloud about the centre of
Centaurus A. In our opinion, this possibility is unlikely.  The overall morphology
and kinematics of the outer \HI\ clouds as observed by SGHK suggest that the
outer \HI\ forms one single, coherent structure of about 30 kpc in length,
half encircling the galaxy from the NE to the SW. The data of SGHK show that
the velocities of this outer \HI\ ring vary smoothly over its entire length of
30 kpc. Hence, SGHK interprete these outer \HI\ structures as forming a
partial ring around Centaurus A that rotates with uniform velocity. A sudden
velocity upturn occurring only at the southern tip is not consistent with a
ring structure 30 kpc in size with regular kinematics.

Even if the outer \HI\ clouds do not form a relatively settled structure, but
instead form a less settled tidal tail-like feature, the anomalous velocities
are not consistent. In general, the velocities along tidal arms vary fairly
smoothly (e.g.\ Hibbard 2003). Velocity reversals are observed, but these are
often due to projection effects and follow naturally from the overall spatial
wrapping of the tidal feature. The reversal observed here in Centaurus A is
different in character since the overall morphology and kinematics of the
outer \HI\ structure does not suggest that wrapping is occurring. In other
cases, velocity reversals are observed near optically bright regions that
could be small objects forming out of the tidal material (i.e.\ Tidal Dwarf
Galaxies, Hibbard 2003, Weilbacher et al.\ 2003).  However, these regions are
always readily visible in the optical as bright star forming``knots'' in the
tidal arm. The \HI\ cloud in Centaurus A discussed here completely lacks such
an optical counterpart.

The location and the kinematics of anomalous \HI\ suggest a different
interpretation, namely that a connection exists between this anomalous gas on
the one hand and the young stars and ionised gas on the other hand.  As is
illustrated in Fig.\ 4, given the location and orientation of the radio jet,
it would hit the \HI\ cloud at the S edge of the \HI\ cloud.  The young stars
and the ionised gas are found downstream from the location with the anomalous
\HI, as one would expect in a jet-cloud interaction scenario.  An important
additional observational fact is that the velocities of the anomalous \HI\
range from 300\kms\ to 430\kms. This coincides quite closely with the velocity
range of the bulk of the ionised gas just E of the \HI\ cloud (Morganti et
al.\ 1991; Graham 1998). The match between the \HI\ and the other features
observed therefore does not only exist in space (which could in principle be a
projection effect), but also in velocity.

\subsection{Jet-induced star formation}

Assuming that the anomalous \HI\ is related to the interaction of the \HI\
cloud with the radio jet, the following may be occurring. The outer \HI\
structure as detected by SGHK is rotating about the galaxy and at a particular
point in time a small region of it is rotating into the relatively narrow cone
defined by the radio plasma. It has often been assumed (e.g.\ SGHK, Graham
1998, Mould et al.\ 2000) that the interaction between the jet and the \HI\
cloud would occur at the E edge of the \HI\ cloud since this is where the
ionised gas and the young stars are found. However, given the jet location and
direction, the S tip is a more likely location where this occurs (see Fig.\
4). The interaction causes some gas to be dragged along the jet and be
displaced from the region where the jet interacts with the \HI. This disturbed
gas is ionised by the interaction (and likely also by the energetic photons
coming from the nucleus of Centaurus A; Morganti et al.\ 1991) while the interaction
also causes the kinematics of the ionised gas to be turbulent (as
observed). Subsequently, a fraction of the ionised gas cools down quickly (see
below) and star formation occurs in the cooled fragments. Given that the gas
is dragged along the jet, the young stars are found displaced from the region
where the jet is interacting with the \HI\ cloud.  The oldest young stars
observed have an age of about 15 Myr, while the displacement of these young stars
from the S tip of the \HI\ cloud is about 2 kpc in projection. This implies
an overall displacement velocity of the ionised gas of the order of
130\kms. This appears consistent, considering differences in projection
factors, with the anomalous velocities detected in the \HI\ that are up to 100
\kms.

Observations of the kinematics of the ionised gas could provide confirmation
of the above model. It is natural to assume that the \HI\ follows a
fractal-like distribution in size and density. Clouds with different sizes and
densities will be accelerated differently and a segregation of clouds will
result away from the location of the interaction. At a given time, smaller
and/or less dense clouds will have been accelerated to larger velocities than
larger and/or denser clouds. This means that starting from the S tip and
following the jet direction, superposed on the very turbulent motions, the
kinematics and the physical conditions of the ionised gas must show a systematic
trend. Kinematical information is available only along a few slit positions
and it is difficult to see whether the data confirms this. Two-dimensional
spectroscopy in several emission lines and covering a large fraction of the
outer filament is needed to study this in more detail.

If the above overall description is correct, the jet induced star formation is
fairly inefficient. Mould et al.\ (2000) report a star formation rate for the
region of the order of a few times $10^{-3}$ \msun\ yr$^{-1}$. Assuming that
the star formation rate has been constant, this implies a total mass for the
stars formed over 15 Myr (the age of the young stars) to be of order of a few
times $10^4$ \msun. The amount of \HI\ showing the anomalous velocities is
about $1\times 10^6$ \msun. Thus, unless the current rate by which the \HI\ is
stripped from the cloud is much higher than in the past, this appears to imply
that the efficiency of converting the  gas  stripped of the cloud into
stars is at most a few percent.

\subsection{Comparison with models}

The only detailed theoretical model of the interaction of the radio plasma
with the environment available for Centaurus~A is the model of the NML by
Saxton et al.\ (2001). Instead of assuming that the NML is a standard radio
lobe, they postulate that it is a buoyant bubble of plasma deposited by an
intermittently active jet. In their model, the ``large-scale jet'' that may be
interacting with the \HI\ cloud is not a classic radio jet but is the trunk of
material trailing the buoyant bubble. They explicitely discuss the interaction
of the trunk with the \HI\ cloud and postulate that the star formation is
triggered by shocks due to the passage of the buoyant material.  Many of the
morphological and kinematical features of the ionised gas can be explained by
this model, as well as the age of the young stars. The flow velocities in this
model are very similar to the flow velocity derived from the displacement of
the young stars from the location of the interaction, giving
further support to this model. Although they do not predict their magnitude,
the fact that the velocities of the anomalous \HI\ are very similar to the
flow velocity indicates that the interaction between the \HI\ cloud and the
buoyant trunk is indeed occurring at the velocities of their model.

A few other models have recently been published regarding the properties of
jet-induced star formation, although they cover a part of parameter space,
e.g.\ shock velocity, that is not directly applicable to Centaurus A. Mellema et
al.\ (2002) model jet-cloud interactions where they postulate that the cocoon
of a fast jet drives a fast shock ($\sim$3500\kms) into ambient clouds of
density 10 cm$^{-3}$. Fragile et al.\ (2004) give similar models with shock
velocities above 1000\kms\ and apply their models to Minkowsky's object.  The
interesting finding in these models is that despite the high velocities
involved, most of the gas cools on a very short timescale (of order $10^2$ yr)
by radiative cooling and forms dense, cool fragments with temperatures below
100 K. Although the conditions are very different from the case of
Centaurus~A, where the shock velocities are most likely much smaller, it
suggests that it is not surprising to see {\sl neutral} gas with anomalous
kinematics. When applying their model to Cygnus~A and trying to reproduce the
star formation rates observed, Mellema et al.\ have to assume an efficiency
for the jet-induced star formation rate of 1\%, while Fragile et al.\ 
have to assume a value of 0.1\% for Minkowsky's object. Again, the conditions
in these models are different from the situation in Centaurus~A, but it is
interesting to see that these values are not too different from our rough
estimate of the star formation efficiency in Centaurus~A.

A possible problem with models of the kind of Mellema et al.\ and Fragile et
al.\ is that it is not clear whether the outflow velocities for the \HI\
predicted by the models are large enough. The fact that the gas very quickly
forms dense clumps means that it is difficult to accelerate them to high
velocities.  In the model of Mellema et al., the fast shock results in
fragments that move with outflow velocities between 90 and 500\kms, with the
densest clumps having the lowest velocity.  The dispersion between that star
associations is about 80\kms. In the model by Fragile et al., the
cold, dense material are hardly accelerated at all.  Given that it is likely
that the shock velocities in Centaurus~A are much smaller that assumed in the
models by Mellema et al.\ and Fragile et al., it remains to be seen whether
they can generate anomalous velocities of the magnitude observed in
Centaurus~A.

\section{Summary}

We have reported the discovery of anomalous velocities of about 100\kms\ in
the \HI\ cloud situated about 15 kpc from the centre of Centaurus~A.  These
anomalous velocities are very localised in the southern tip of the cloud and
we suggest that they are related to the interaction between the radio jet and
the neutral hydrogen.  The cooling of the gas is then responsible for the
formation of the young stars observed downstream. From the displacement of the
young stars from the location of the anomalous velocities we derive a flow
velocity of about 130\kms. The jet induced star formation appears to be fairly
inefficient, of the order of few percent.

The results presented here give further support to the idea that star
formation can be produced via interaction of a radio jet with the ISM.

\begin{acknowledgements}

This work is based on observations with the Australia Telescope Compact Array
(ATCA), which is operated by the CSIRO Australia Telescope National
Facility. We thank David Malin for supplying the optical image used in
Fig.\ 1 and Marina Rejkuba for providing the optical image used
in Fig.\ 4.
\end{acknowledgements}


\begin{thebibliography}{}


\bibitem[]{} Blanco, V.M., Graham, J.A., Lasker, B.M., Osmer, P. 1975, 
ApJ, 198, L3
\bibitem[]{}  Briggs, D. 1995, Ph.D. thesis, New Mexico Inst. Mining Tech. 
\bibitem[]{} Charmandaris, V., Combes, F. \& van der Hulst, J.M. 2000, A\&A, 356, L1
\bibitem[]{} Dey, A., van Breugel, W.,  Vacca, W.D., Antonucci, R. 1997, ApJ, 490, 698
\bibitem[]{} Fassett, C.I. \& Graham, J.A. 2000, ApJ, 538, 594
\bibitem[]{} Fragile, P.C., Murray, S., Anninos, P., van Breugel, W. 2004, ApJ,
604, 74
\bibitem[]{} Graham, J.A. 1998, ApJ, 502, 245
\bibitem[]{} Hibbard, J.E., 2003,  in:  
IAU Symp. 217, {\em Recycling Intergalactic and Interstellar
Matter}, eds. P.-A. Duc, J. Braine, and E. Brinks,  p. 82
\bibitem[]{}  Hui, X., Ford, H.C., Ciadullo, R., Jacoby, G.H. 1993, ApJ,  414, 463
\bibitem[]{} Mellema, G., Kurk, J.D., R\"ottgering, H.J.A. 2002, A\&A, 395, L13
\bibitem[]{} Morganti, R., Robinson, A., Fosbury, R.A.E., di Serego Alighieri,
S., Tadhunter, C.N., Malin, D.F. 1991, MNRAS, 249, 91
\bibitem[]{} Morganti, R., Killeen, N.E.B., Ekers, R.D., Oosterloo, 
T.A. 1999, MNRAS, 307, 750
\bibitem[]{} Mould, J.R. et al. 2000, ApJ,  536, 266
\bibitem[]{} Neff, S.G., Schiminovich, D., Martin, C.D., GALEX Science Team 
2003, AAS, 103, 9607
\bibitem[]{} Rees, M.J. 1989, MNRAS, 239, 1
\bibitem[]{} Rejkuba, M., Minniti, D., Courbin, F., Silva, D.R. 2002, ApJ, 564, 688

\bibitem[]{} Sault, R.J., Teuben, P.J. \& Wright, M.C.H. 1995, in Astronomical
  Data Analysis Software and Systems IV, R. Shaw, H.E. Payne and J.J.E. Hayes,
  eds, Astronomical Society of the Pacific Conference Series, 77, p. 433
\bibitem[]{} Saxton, C.J., Sutherland, R.S, Bicknell, G.V. 2001, ApJ, 563, 103
\bibitem[]{} Schiminovich, D., van Gorkom, J.H., van der Hulst, J.M., 
Kasow, S. 1994, ApJ, 423, L101  (SGHK)
\bibitem[]{} Stickel, M., van der Hulst, J.M., van Gorkom, J.H., 
Schiminovich, D., Carilli, C.L. 2004, A\&A, 415, 95
\bibitem[]{} Sutherland, R.S., Bicknell, G.V., Dopita, M.A.  1993, ApJ, 414, 510
\bibitem[]{} van Breugel, W., Filippenko, A.V., Heckman, T.M., Miley, G.K. 1985,
ApJ, 293, 83
\bibitem[]{} van Breugel, W., Dey, A. 1993, ApJ, 414, 563
\bibitem[]{} van Breugel, W. 2000, Proc. SPIE Vol. 4005, p. 83-94
(astro-ph/0006238)
\bibitem[]{} van Breugel, W., Fragile, P.C., Anninos, P., Murray, S.  2003, in:  
IAU Symp. 217, {\em Recycling Intergalactic and Interstellar
Matter}, eds. P.-A. Duc, J. Braine, and E. Brinks,  p. 29
(astro-ph/0312282)
\bibitem[]{} Weilbacher, P.M., Fritze-von Alvensleben, U., Duc, P.-A.
2003, in:  
IAU Symp. 217, {\em Recycling Intergalactic and Interstellar
Matter}, eds. P.-A. Duc, J. Braine, and E. Brinks, p. 16

\end{thebibliography}
\end{document}